\documentstyle[12pt,epsf,a4]{article}
\begin{document}
\baselineskip 0.8 true cm

\def\gms{$G_{Ms}$}
\def\gmp{$G_{Mp}$}
\def\gmn{$G_{Mn}$}
\def\ges{$G_{Es}$}
\def\gep{$G_{Ep}$}
\def\gen{$G_{En}$}
\begin{center}
{ \Large \bf
 Recent results on hadron form factors.}
\vspace*{.5 true cm}

{\rm\large Egle Tomasi-Gustafsson and Michail P. Rekalo\footnote{ Permanent address:\it National Science Center KFTI, 310108 Kharkov, Ukraine}}

{\it DAPNIA/SPhN, CEA/Saclay, 91191 Gif-sur-Yvette Cedex,
France}

\end{center}

\begin{center}
{\Large\bf Abstract}
\end{center}
We discuss the recent data on the electric proton form factor, obtained at JLab, which definitely show a spectacular deviation from the commonly assumed dipole behavior. We discuss the implication of these results
on the deuteron structure and on the neutron electric form factor: at relatively large $Q^2$ a revision of the deuteron models may be required, and the neutron electric form factor might become even larger than the proton electric form factor.

\section{Introduction}
The complex structure of hadrons can be described in a convenient way in terms
of form factors. In a parity conserving and time invariant theory, each particle
of spin $S$ can be described in terms of $2S+1$ elastic electromagnetic form
factors, which are
real functions in the space-like region and complex functions in the time-like
region of momentum transfer square.

The (elastic or inelastic) electron-hadron scattering  is the traditional way
to determine the form factors, and it allows a
direct comparison with the theory \cite{Dr61}. The precise measurement of these form factors requires polarization experiments
(except for spin zero particles, like pions or kaons). Recent measurements at the Jefferson
Laboratory essentially improved the experimental data concerning the elastic
form factors of protons and deuterons up to relatively large values of momentum
transfer \cite{Jo00,Al99,Ab00}. The study of the internal structure of light hadrons is typically one of the main
investigations which can be carried on with a high intensity, high duty cycle electron machine as Jefferson Laboratory (JLab). A large program is under way to measure the form factors of the neutron and of the first nucleon resonances \cite{Bu95,Zhu01,Madey}.
Information on the internal structure of the nucleon is
also provided by the annihilation channels as 
$e^+ +e^-\leftrightarrow p+ \overline{p}$ \cite{ETG01}. 

We will discuss the implications of these new results and the related opportunities for the Nuclotron, due to its unique capability to accelerate polarized deuterons.

\section{Proton electromagnetic form factors}

The elastic $ep$ cross section, in one-photon exchange approximation, can be  written as a function of the electric $G_{Ep}$ and magnetic $G_{Mp}$ proton form factors \cite{Rosenbluth}:
$$\displaystyle\frac{d\sigma}{d\Omega}=
\left (\displaystyle\frac{d\sigma}{d\Omega}\right)_{Mott}\left [ \displaystyle\frac {G_{Ep}^2+\tau G_{Mp}^2}{1+\tau} +2\tau 
G_{Mp}^2\tan^2\displaystyle\frac{\theta_e}{2}\right ],~ \tau=\displaystyle\frac{Q^2}{4M^2} $$
with
$$
\left (\displaystyle\frac{d\sigma}{d\Omega}\right)_{Mott}=
\frac{\alpha^2~\cos^2(\theta_e/2)E'}{4E^3\sin^4(\theta_e/2)} \mbox{~and~} 
E'=\displaystyle\frac {E}{ 1+2\displaystyle\frac {E}{M}\sin^2(\theta_e/2)}$$
where $M$ is the proton mass, $E$ is the energy of the incident electron,
$E'$ and $\theta_e$ are the energy and scattering angle of the outgoing electron,
$\alpha$ is the fine structure constant. The momentum of the virtual photon is $Q^2=4EE'\sin^2(\theta_e/2)$ and it is positive in the space-like region. Measurements of the cross section for the same $Q^2$ and different angles allow, in principle, to determine $G_{Ep}^2$ and $G_{Mp}^2$. However, at larger $Q^2$, it is very difficult to disentangle the electric form factor with high precision, as its contribution dies out (it is $\simeq 8\%$ at $Q^2=9$ GeV$^2$). Therefore a measurement of the differential cross section allows to determine precisely only $|G_{Mp}|$. More than forty years ago it was shown in \cite{Ak76} that the polarized cross section (with longitudinally polarized electrons on a polarized proton target, or with measurement of the recoil proton polarization) contains a term proportional to $G_{Ep}G_{Mp}$ and it was suggested for the first time a very sensitive method for the determination of $G_{Ep}$. Only recently this method could be applied, as it needs high high intensity polarized beams, large solid angle spectrometers and advanced techniques of polarimetry \cite{Jo00,Mi98}. 

\subsection{The Recoil Polarization Method}

The polarization method was first discussed by Akhiezer and
Rekalo \cite{Ak76}, then Dombey \cite{Do69}, 
and later by Arnold, Carlson and Gross \cite{Ar81}. The idea is to measure the
transferred longitudinal and sideways polarizations, $P_{\ell }$
and $P_{t}$ in $\vec e p\rightarrow e\vec p$, of the
recoiling proton with a polarimeter.

Assuming one photon exchange there is no out-of-plane polarization transfer
($P_n$=0), and for a 100 $\%$ longitudinally polarized beam, $P_{\ell }$
and $P_{t}$ are:
\begin{eqnarray}
I_{0}P_{\ell }&=&\frac{E_{e}+E_{e^{\prime }}}{M}\sqrt{\tau (1+\tau )}
G_{Mp}^{2}\tan ^{2}\frac{\theta _{e}}{2} \label{eq:pl}\\
I_{0}P_{t}&=&-2\sqrt{\tau (1+\tau )}G_{Ep}G_{Mp}\tan \frac{\theta
_{e}}{2}  \label{eq:pt}
\end{eqnarray}
where
\begin{equation}
I_{0}=G_{Ep}^{2}(Q^{2})+\tau G_{Mp}^{2}(Q^{2})\left [1+2(1+\tau )\tan ^{2}\frac{
\theta _{e}}{2}\right ].
\end{equation}

One of the advantages of the polarization method is that the interesting 
observable, $P_{t}$, is an interference term; thus even
a small $G_{Ep}$ will get amplified by a large $G_{Mp}$.

The proton polarization is measured by a polarimeter installed at the focal plane of a spectrometer. The protons, issued from a primary target, are momentum analyzed in the spectrometer and undergo a second scattering in a thick analyzer (usually carbon or polyethylene). The outgoing trajectories of the charged particles are reconstructed in a detection with $2\pi$ azimuthal angular acceptance. For each polar ($\vartheta $) and 
azimuthal ($\varphi$) angle and for each helicity state $\pm$ of the beam, the event distribution after scattering in the analyzer can be written as:
\begin{eqnarray}
N_{p}^{\pm}(\vartheta ,\varphi )=N_{p}^{\pm}
\epsilon (\vartheta )\left [1\pm A_{c}(\vartheta
)\left (P_{t}^{fpp}\sin \varphi - P_{n}^{fpp}\cos \varphi \right )\right ], 
\label{eq:nphi}
\end{eqnarray}
where $N_{p}$ is the number of protons
incident on the polarimeter, 
$\epsilon (\vartheta )$ is the differential efficiency, and $A_{c}(\vartheta )$
the analyzing power of the analyzer;
$P_{t}^{fpp}$ and $P_{n}^{fpp}$ are the transverse and normal
components of the polarization at the analyzer. 

In the magnetic elements of a spectrometer spin precession occurs and the polarizations at the primary target can be exactly calculated knowing the spin transfer matrix.
For each $Q^{2}$ a single measurement of the
azimuthal angular distribution of the protons diffused in a secondary target 
determines simultaneously both $P_{\ell }$ and $P_{t}$.

The ratio $G_{Ep}/G_{Mp}$ can then be obtained directly from the ratio:
\begin{equation}
\displaystyle\frac{ G_{Ep}}{G_{Mp}}=-\displaystyle\frac{(E_{e}+E_{e^{\prime }})}{2M}\tan
\left (\frac{\theta _{e}}{2}\right )
\end{equation}
Moreover, knowing the beam polarization (which is 
measured independently) and  $G_{Mp}$, one can derive the values of the
polarimeter analyzing powers for each of the proton energies.

\subsection{Results}

The recoil polarization method has been 
successfully used to measure the ratio $G_{Ep}/G_{Mp}$ in Hall A at JLab, in two steps: firstly a measurement up to  $Q^{2}$ =3.5
GeV$^{2}$  \cite{Jo00} and, later, up to $Q^{2}$ =5.6
GeV$^{2}$ \cite{E99-007}.
In both experiments the focal plane polarimeter (FPP) was installed in the focal plane of the 4 GeV/c high resolution magnetic
spectrometer (HRS).   Two changes were made to
allow continuation to larger $Q^2$: first the 50 cm of carbon analyzer was replaced 
with 100 cm of CH2, to take advantage of the larger analyzing power of the
hydrogen. Second, the electron was detected in lead-glass Cerenkov detector 
array with 3.5 $m^2$ frontal area, replacing the second HRS previously used.
This technique allows for matching the solid angle of the hadron to the one 
of the electron increasing the counting rate by a factor up to six.

\begin{figure}[!]
\vspace*{-2truecm}
\begin{center}
\mbox{\epsfxsize=11.cm\leavevmode\epsffile{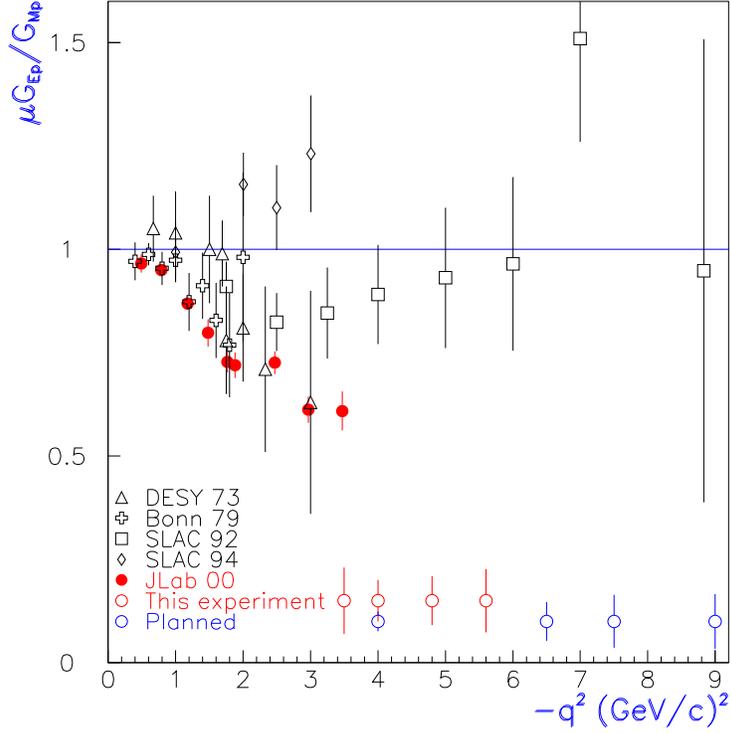}}
\vspace*{-0.5truecm}

\caption{Ratio $\mu G_{Ep}/ G_{Mp}$ as a function of $Q^2$. The solid circles  are the results from \protect\cite{Jo00}, compared to previous data based on Rosenbluth separation \protect\cite{Ar86} and refs. herein. On the bottom shown the 
Q$^2$ values and the expected error bars from \cite{E99-007} and \cite{00111}}
\label{fig:gep}
\end{center}
\end{figure}

These experiments showed that the ratio of the two elastic form factors
of the proton, electric and magnetic, $G_{Ep}/ G_{Mp}$, decreases monotically
with increasing four momentum squared (Fig. \ref{fig:gep}).

The magnetic form factor of the proton, $G_{Mp}$, is
known to remain within a few percent of $\mu_pG_D$, over this range of $Q^2$,
where $G_D$ is the dipole form factor: 
$G_D=\left [1+{Q^2}/0.71 ~\mbox{GeV}^2\right ]^{-2}$. It follows that $G_{Ep}$ decreases significantly faster than
$G_D$ and $G_{Mp}$ in the measured interval. In the Breit system, form factors are related to the Fourier transform of the charge and magnetic moment
distribution and the dipole approximation results from an
exponential distribution. The new results indicate that
the electric charge distribution in the proton extends to larger distances than
the magnetization distribution. The vector dominance model predicts a decrease of the
$G_{Ep}/ G_{Mp}$ ratio with increasing $Q^2$, and the same prediction comes from various relativistic constituent quark models, but these models depend on a large number of parameters, which are not constrained by other experimental data at this time. At a few GeV incident beam energy, one expects to 'see' the manifestation of
quark degrees of freedom. Perturbative QCD gives definite rules about the
scaling behavior of the form factors and about helicity conservation, at very large $Q^2$. Other approaches, as lattice QCD calculations are still in a very early stage and formalisms based on  generalized parton distributions lack predictive power.

We can fairly summarize the situation stating that there is 
not yet a coherent picture of the nucleon structure, in framework of any of these models. Moreover one aims to a global interpretation
of form factors data, for proton and for neutron, in space-like and in time-like region, as well.

On the other hand, as we will show in next sections, one can already foresee major consequences of these data. The models 
describing the light nuclei structure usually assume a dipole behavior for
$G_{Ep}$ and a vanishing or negligible neutron electric form factor $G_{En}$. Elastic electron-deuteron scattering is sensitive to the isoscalar form
factor, $G_{Es}=G_{Ep}+G_{En}$. If $G_{Ep}$ turns out to be smaller than 
previously assumed, this has to be compensated either by $G_{En}$ or by other 
ingredients used in  deuteron calculations (which are parametrized or 
adjusted on the data).

The same
technique is proposed to measure the $G_{Ep}/G_{Mp}$ ratio to $Q^{2}$ = 9.0 GeV$^{2}$
in Hall C at JLab \cite{00111}. The future experiment will take advantage
of the higher momentum capability of the spectrometer (HMS) of Hall C. A new polarimeter will be built, and a new calorimeter with finer granularity
will be assembled for this purpose. The experiment will extend the data up to $Q^2\simeq $ 9 GeV$^2$, with the
current JLab maximum electron energy of 6 GeV. With the same set up in Hall C 
an extension up to at least $Q^2\simeq $ 11.5 GeV$^2$ will become possible
after the proposed upgrade of the accelerator to 12 GeV.

\subsection{Opportunities for the Nuclotron}

Polarization experiments are, in general, very lengthy and time consuming. Therefore
a thorough optimization of the characteristics of the polarimeter is desired.
This requires a careful study of the analyzing reaction, which has to have large cross section and large analyzing powers.
The crucial feature of the polarimeter is its figure of
merit ${\cal F}$, defined as ${\cal F}=\epsilon A_y^2$, where $\epsilon$ is the useful fraction of events scattered in the analyzer, and $A_y^2$ is the
analyzing power squared. Adding hydrogen
to the analyzer increases the analyzing power. The knowledge of $A_y$ to large momenta is highly desirable for preparing the future experiment. A calibration at different momenta, ranging from 3.7 to 5.4 GeV/c has been performed at the JINR-LHE accelerator complex, after installing the POMME polarimeter, transported from Saclay \cite{dubnacalib}. Starting from a polarized deuteron beam of
9 GeV/c incident on a primary target (C or LH$_2$),
one can magnetically select the breakup
protons with definite momentum; their polarization is known from previous breakup studies at Saclay and 
Dubna \cite{punjabi,kuehn}. The Nuclotron can play a major role in the optimization of the characteristics of the polarimeter, like efficiency and figure of merit.

\section{The deuteron electromagnetic form factors}

The measurement of the differential cross section of elastic $ed-$scattering,
for a fixed value of $Q^2$,
at
different scattering angles, allows to determine the structure functions
$A(Q^2)$
and $B(Q^2)$:
$$\displaystyle\frac{d\sigma}{d\Omega}=
\left (\displaystyle\frac{d\sigma}{d\Omega}\right)_{Mott}\cdot
{\cal S},~~{\cal S}= A(Q^2)+B(Q^2) ~\tan^2(\theta_e/2).
$$

Being the deuteron a spin 1 particle, $A$ and $B$ can be expressed in terms of the three form
factors, $G_c$ 
(electric), $G_m$ (magnetic) and $G_q$ (quadrupole) as:
\begin{equation}
A(Q^2) =G_c^2(Q^2)+ \frac{8}{9} \tau_d^2 G_Q^2(Q^2)+\frac{2}{3} \tau_d
G_m^2(Q^2),~~B(Q^2) =
\frac{4}{3} (1+\tau_d) \tau _dG_m^2(Q^2),~~\tau_d=\frac{Q^2}{4M_d^2},
\label{eq:aq}
\end{equation}
where $M_d$ is the deuteron mass. The measurement of the differential cross section is not sufficient to disentangle the three form factors. In case of unpolarized beam and target, the outgoing deuteron is tensorially polarized and the components of the tensor
polarization give useful combinations of form factors, in particular $t_{20}$: 
$$
t_{20}=-\displaystyle\frac{1}{\sqrt{2}{\cal S}}\left \{
\displaystyle\frac{8}{3}\tau_d G_c G_q + \displaystyle\frac{8}{9} \tau_d^2
G_Q^2+\frac{1}{3} \tau_d
\left [1+2(1+\tau_d)\tan^2 (\theta_e/2)\right ]G_m^2\right \},
$$
At the Jefferson Laboratory (JLab), the elastic $ed-$ cross section has been
recently precisely measured up to  $Q^2\simeq 6$ GeV$^{2}$,
\cite{Al99} and $t_{20}$ up to $Q^2=1.9$ GeV$^2$  \cite{Ab00}.

According to \cite{Al99}, the cross sections seems to scale as $(Q^2)^{-10}$, as
previously pointed out \cite{Ar86}, and predicted by pQCD. The authors of \cite{Al99} suggest that the data about the structure
function $A(Q^2)$ in $ed$ elastic scattering, in the range  $2\le Q^2\le$ 6 GeV$^2$ are a good indication of the validity of the predictions of pQCD.
However, from the $t_{20}$ data, it appears that the pQCD
limit is not
yet reached, and the data follow the trend suggested by impulse approximation (IA). On the other
hand, it is not possible, from these data, to constrain definitely different
models or determine unambiguosly the corrections to IA (for a detailed comparison with some theoretical models, see, for example, \cite{Al99,Ab00}).  Following \cite{Br76}, one can define a generalized deuteron form factor, $F_D(Q^2)$, $F_D(Q^2)=\sqrt{A(Q^2)}$, and a reduced deuteron form factor $f_D(Q^2)$:
\begin{equation}
f_D(Q^2)=\displaystyle\frac{F_D(Q^2)}{F_N^2(Q^2/4)},
\label{eq:eq1}
\end{equation}
where $F_N$ is the nucleon electromagnetic form factor. The $(Q^2)$-behavior of 
$f_D(Q^2)$ (at large $Q^2$) can be predicted in the framework of pQCD, according to:
\begin{equation}
f_D(Q^2)=N\displaystyle\frac{\alpha_s(Q^2)}{Q^2}\left ( ln \displaystyle\frac{Q^2}{\Lambda^2}\right)^{-\Gamma},
\label{eq:eq2}
\end{equation}
where N is the normalization factor (which can not be calcuated by QCD), $\alpha_s$ is the running QCD strong interaction coupling constant, $\Lambda$ is the scale QCD parameter, and $\Gamma$ is determined by the leading anomalous dimension, here $\Gamma=-8/145$. In \cite{Al99} it was shown that the QCD prediction (\ref{eq:eq2}), which applies to asymptotic momentum transfer, is working well already for $Q^2\ge 2$ GeV$^2$, with a plausible value of the parameter
$\Lambda \simeq$ 100 MeV.

In ref. \cite{Br76}, 
the value of the nucleon form factor $F_N$ has been parametrized in dipole form: \begin{equation}
F_N(Q^2)= G_D=\displaystyle\frac{1}{(1+Q^2/0.71\mbox{~GeV}^2)^2}
\label{eq:eq4}
\end{equation}
and it was not rigorously identified as magnetic or electric, proton or neutron.

From  quark counting rules considerations, the dipole form of the nucleon form factors has been taken until recently as universal, consistent with the experimental data for three of the four nucleon form factors, $G_{Mn}$, $G_{Mp}$, and $G_{Ep}$. The fourth, $G_{En}$, was assumed negligible in the discussed region of $Q^2$.

Before the JLab measurement \cite{Jo00}, the experimental data about $eN$-scattering, based on Rosenbluth separation \cite{Ar86} were consistent with this representation. The data from \cite{Jo00} can be fitted by:
\begin{equation}
G_{Ep}(Q^2)= \displaystyle\frac{1}{(1+Q^2/0.71\mbox{~GeV}^2)^2}\displaystyle\frac{1}{(1+Q^2/4.8\mbox{~GeV}^2)},
\label{eq:eq5}
\end{equation}
where the second factor explicitely shows the deviation from the dipole form.
The data can still be described by a dipole form, but changing the well known mass parameter 
$m^2_D=0.71$ GeV$^2$ to a smaller value: $m^2_D=0.6$ GeV$^2$. This parametrization may seem preferable, because consistent with the pQCD counting rules, but the best fit value of $m^2_D$ is in contradiction with the nice relation between the $Q^2$-behavior of pion and nucleon form factors, derived in \cite{Br76}.

\begin{figure}[!]
\begin{center}
\vspace*{-3true cm}
\hspace*{3true cm}\mbox{\epsfxsize=12.cm\leavevmode \epsffile{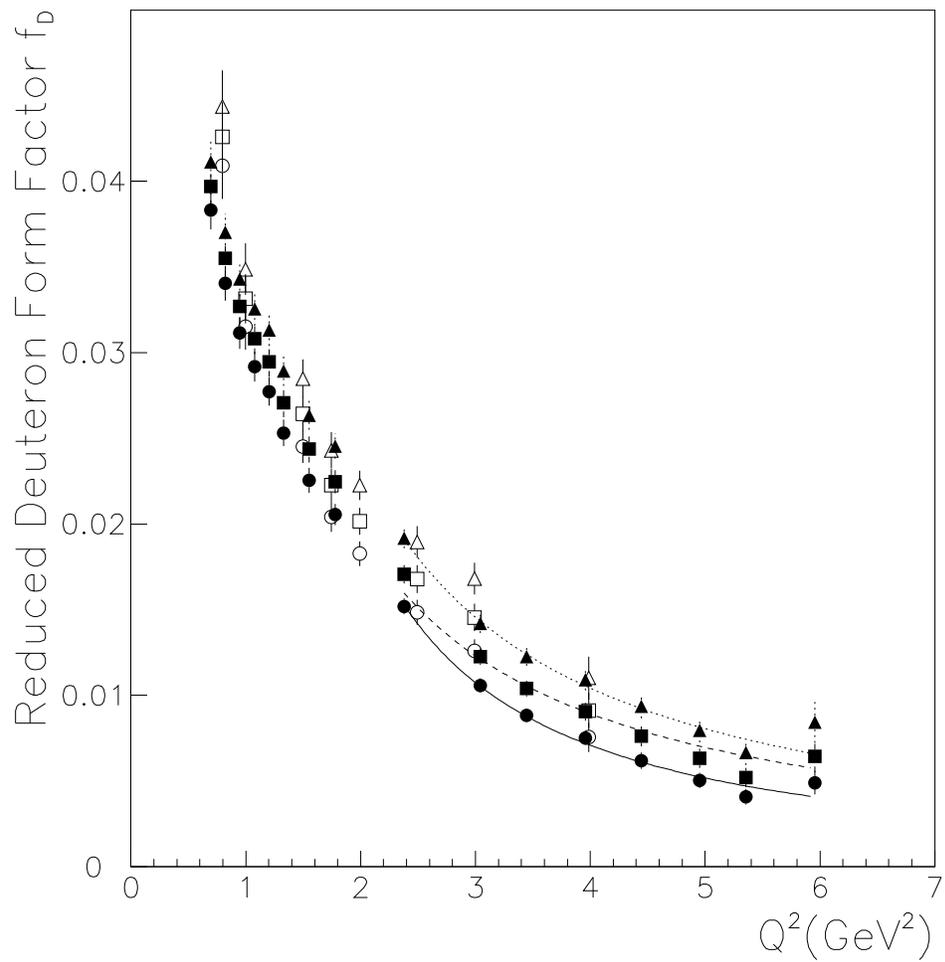}}
\vspace*{-2truecm}
\end{center}
\caption{Data set corresponding to the reduced deuteron form factor for different choices of the generalized nucleon form factor: triangles (case 1), squares (case 2) circles (case 3). Solid symbols are from \protect\cite{Al99}, open symbols from \protect\cite{Ar86}.}
\label{fig:fig1}
\end{figure}

In Fig. (\ref{fig:fig1}) we show different data sets and best fits, using Eq. (\ref{eq:eq2}), corresponding to the following possibilities:
\begin{enumerate}
\item We replace, in Eq. (\ref{eq:eq1}), $F_N$ by the  fit (\ref{eq:eq5}) of new data on the proton electric form factor, $G_{Ep}$: $$f_D(Q^2)=\displaystyle\frac{F_D(Q^2)}{G_{Ep}^2(Q^2/4)}.$$ This yields to the data set represented by triangles and to the fit reported as a dashed line (case 1).
\item  We replace, in Eq. (\ref{eq:eq1}), $F_N^2$ by the product of $F_N$, Eq. (\ref{eq:eq4}), and $G_{Ep}$ from (\ref{eq:eq5}): $$f_D(Q^2)=\displaystyle\frac{F_D(Q^2)}{F_N(Q^2/4) G_{Ep}(Q^2/4)}.$$ The $f_D$ data are shown as squares and the best fit by dotted line (case 2).
\item We show, for comparison, the previous results of Ref. \cite{Al99}, using Eq. (\ref{eq:eq4}). The data are represented by circles and the fit by the solid line (case 3).
\end{enumerate}

In all these three cases, instead of normalizing the model to the data at $Q^2= 4$ GeV$^2$, as in Ref. \cite{Al99}, we have fitted the data beyond $Q^2= 2$ GeV$^2$, according to Eq. (\ref{eq:eq2}),  with two free parameters, a global normalization $N$ and $\Lambda$.  We found that even a relatively small change in nucleon form factors, causes a large instability in the value of $\Lambda$. Note that $f_D$ has logarythmic (i.e; relatively weak) dependence on $\Lambda$ (Eq. (\ref{eq:eq2})).  For case 3, we obtain a different value compared to Ref. \cite{Al99}, due to the different normalization procedure. 

The best fit parameters are reported in Table 1. The values which can be obtained for $\Lambda$ may differ by an order of magnitude, according to the choice of the nucleon form factors.
A similar situation occurs if we use Dirac and Pauli form factors, $F_1$ and $F_2$, instead of the Sachs form factors $G_E$ and $G_M$.

\begin{center}

\begin{tabular}{|c|c|c|}
\hline\hline
Case&N & $\Lambda [GeV]$\\
\hline\hline
 (1)& 225 $\pm$ 65& .101  $\pm$ .089\\
 (2)& 263 $\pm$ 35& .032$\pm$ .016\\
 (3)& 61 $\pm$ 20 & .648$\pm$ .228\\
 \hline\hline
\end{tabular}
\vspace*{.2true cm}

{\small{\bf Table 1} Values of the fit parameters, corresponding to Fig. \ref{fig:fig1}. See text. }
\end{center}

\vspace*{.2true cm}

\begin{figure}[!]
\vspace*{-2true cm}
\begin{center}
\mbox{\epsfxsize=9.cm\leavevmode\epsffile{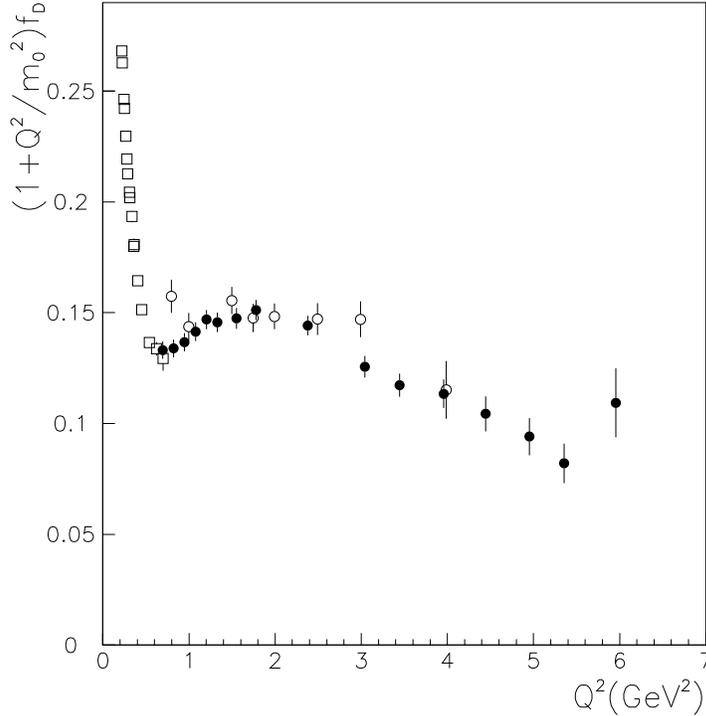}}
\vspace*{-2true cm}
\caption{Data set corresponding to the reduced deuteron form factors multiplied by $(1+Q^2/m_0^2)$. Solid circles from \protect\cite{Al99}, open circles are from \protect\cite{Ar86}, open squares from \protect\cite{Pl90}.}
\label{fig:fig2}
\end{center}
\end{figure}

In  \cite{Br76} another interesting prediction, concerning the scaling behavior of the reduced deuteron form factor was done:
\begin{equation}
\left (1+\displaystyle\frac{Q^2}{m_0^2}\right )f_D(Q^2)\simeq const,
\label{eq:eq3}
\end{equation}
where $m_0^2=0.28$ GeV$^2$ is a parameter related to the pion form factor. The same data from \cite{Al99}, if plotted in the representation of the reduced deuteron form factors, should illustrate the $Q^2$-independence of this product. This result is consisten by the previous $A(Q^2)$ data \cite{Ar86}, in the limit of their accuracy, but not with the new data about $A(Q^2)$ \cite{Al99} (Fig. (\ref{fig:fig2})). This is also true for the different choices of the electromagnetic nucleon form factors considered above. This result is quite insensitive to different values of the $m_0$ parameter. 

One should also take into account the fact that the elastic $ed$-scattering is sensitive to the isoscalar combination of the nucleon form factors. So a linear combination of proton and neutron form factors seems more adequate for the parametrization of $F_N$. In the case of dipole parametrization of nuclear electromagnetic form factors, an isoscalar combination will only bring a different value of normalization. But, if one takes $G_{En}\ne$ 0, two other possibilities: $F_N^2=G_{Es}^2$ and $F_N^2=G_{Es}G_{Ms}$ would lead to different results and different values for $\Lambda$.

We discussed above the sensitivity of the reduced deuteron form factor to different choices of nucleon form factors. However, the numerator of Eq. (\ref{eq:eq1}) contains a generalized deuteron form factor, derived from the structure function $A(Q^2)$. It would be more natural to include the electric, quadrupole or magnetic deuteron form factors, $G_E$, $G_Q$, and $G_M$ in the calculation of $f_D$. 

Therefore we conclude that the situation with nucleon and deuteron electromagnetic form factors, at the light of the recent $G_{Ep}$ data, in the intermediate $Q^2$ range is less clear then in case of dipole parametrization for all form factors; the results of \cite{Jo00} open the way to new interpretations in different directions. We must stress that these considerations are essentially  based on the existing experimental information about nucleon form factors.

\section{The neutron electromagnetic form factors}

Having high precision data on the differential cross section for $ed-$ elastic
scattering, and assuming a
reliable model for their description,  one can extract, in principle, the
dependence of the electric neutron form factor \gen\  on the momentum transfer
$Q^2$. Such a procedure  has been carried out in ref. \cite{Pl90}, up to
$Q^2$=0.7 GeV$^2$. It can be extended at higher $Q^2$ \cite{Etg01} using the  elastic
$ed$-scattering data mentioned above and the recent data on the proton electric
form factor \cite{Jo00}. 
The large sensitivity to the nucleon form factors of the models which
describe the light nuclei structure, particularly the deuteron, was already
carefully studied in \cite{Ar80}, and
it was pointed out that the disagreement between the relativistic impulse
approximation and the data could be
significantly reduced if \gen\  were different from zero.

In the non relativistic IA, the deuteron form factors
depend only on the deuteron wave function and on nucleon form factors:
\begin{equation}
G_c=G_{Es}C_E,~~G_q=G_{Es}C_Q,~~G_m=\displaystyle\frac{M_d}{M}\left
(G_{Ms}C_S+\displaystyle\frac{1}{2}G_{Es}C_L\right ),
\label{eq:gc}
\end{equation}
where  \ges=\gep+\gen\  and \gms=\gmp+\gmn\  are the
charge and
magnetic isoscalar nucleon form factors, respectively. The terms
$C_E$, $C_Q$,
$C_S$, and $C_L$ describe the deuteron structure and can be calculated from the
deuteron $S$ and $D$
wave functions, $u(r)$ and $w(r)$ \cite{Ja56} :
$$C_E={\int }_0^{\infty}dr~j_0\left(
\frac{Qr}2\right) \left[ u^2\left( r\right) +w^2( r)
\right], $$
$$C_Q=\frac{3}{\sqrt{2}\eta}{\int }_0^{\infty}dr~j_2\left(
\frac{Qr}2\right) \left[ u( r) -\frac{w( r)}{\sqrt{8}}\right] w(r),  $$
\begin{equation}
C_S={\int }_0^{\infty}dr \left[ u^2( r) -\frac{1}2w^2( r)
\right ]j_0\left( \frac{Qr}{2}\right) +
\frac{1}{2}\left [\sqrt{2}u( r)w(r)+w^2( r)\right ] j_2\left(
\frac{Qr}{2} \right ),
\label{eq:ds}
\end{equation}
$$C_L=\frac{3}{2} \int_0^{\infty}dr~w^2( r)
\left [ j_0 \left ( \frac{Qr}{2} \right )+ j_2 \left ( \frac{Qr}{2}\right )
\right ],
$$
where
$j_0(x)$ and $ j_2( x)$
are the spherical Bessel functions.
The normalization condition is $
{\int }_0^{\infty}dr~\left[ u^2( r)+w^2( r)\right ]=1.$

With the help of expressions (\ref{eq:gc}) and (\ref{eq:ds}), the formula (\ref{eq:aq}) for $A(Q^2)$,
can be inverted into a quadratic equation for \ges. Then \ges\  can be calculated
using the experimental values for $A(Q^2)$, assuming, for the magnetic nucleon
form factors $G_{Mp}$ and $G_{Mn}$
the usual dipole dependence, which is in fair agreement with the existing data up to $Q^2\simeq$ 10 GeV$^2$.

\begin{figure}
\vspace*{-2truecm}
\begin{center}
\mbox{\epsfxsize=14.cm\leavevmode \epsffile{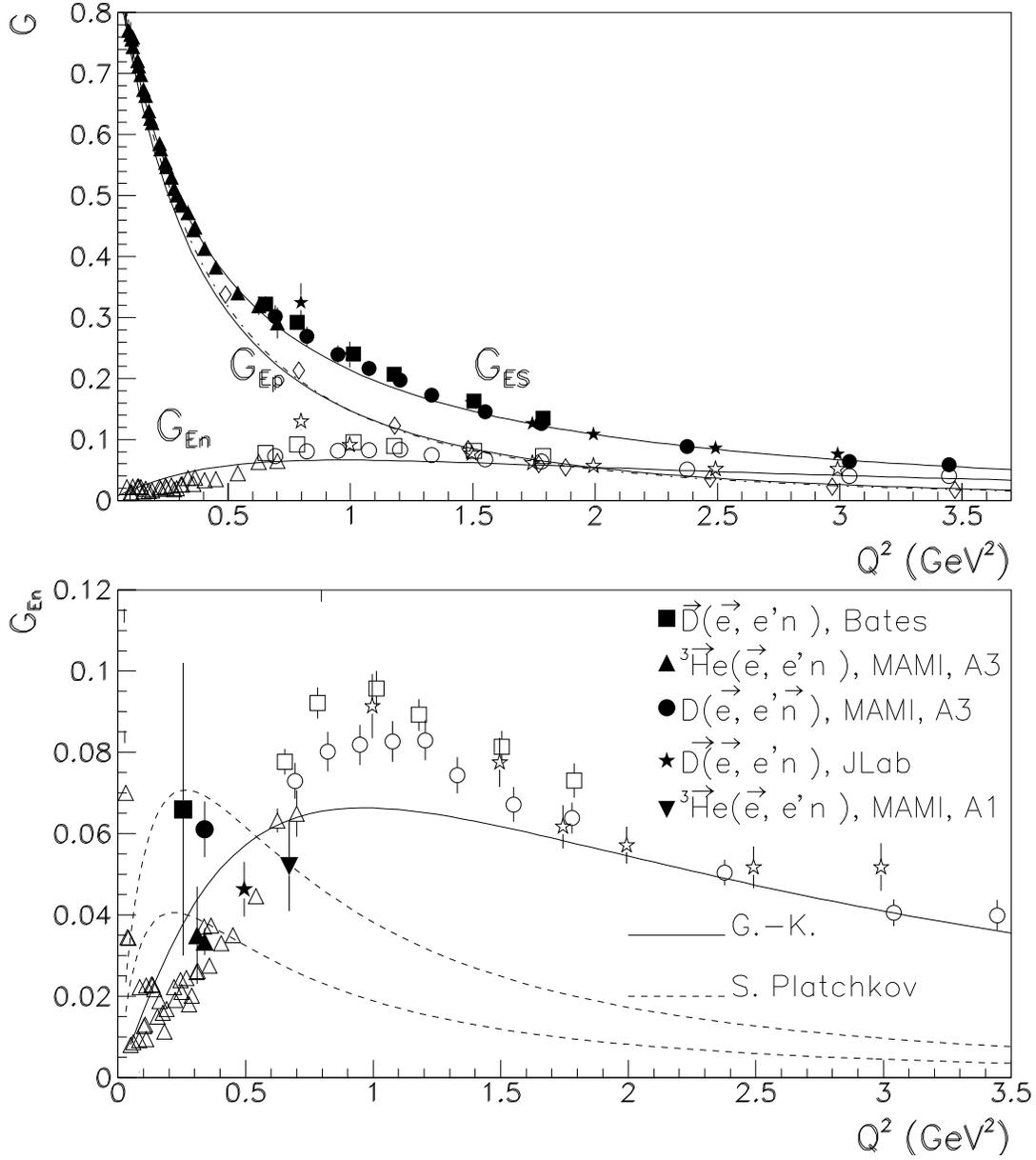}}
\end{center}
\vspace*{-2truecm}
\caption{Top: nucleon  electric form factors as functions of
the momentum transfer $Q^2$. in the framework of IA with Paris potential.
Isoscalar electric  form factors are derived from  the deuteron elastic
scattering data: \protect\cite{Pl90} (solid triangles), \protect\cite{Al99}
(solid circles), \protect\cite{Ab99} (solid squares), and \protect\cite{Ar75}
(solid reversed triangles). The  electric proton form factors from 
\protect\cite{Jo00} are shown as open diamonds. The  electric neutron form
factors, are shown as open symbols. The parametrization
\protect\cite{G-K} is drawn for all three form factors as a solid line. The dashed-dotted line is the
parametrization  for \gep\  from Eq. (\ref{eq:eq5}) and it is hardly distinguishable from \protect\cite{G-K}. Bottom: neutron  electric form factors, in the low $Q^2$ region, compared to 'direct' measurements (solid simbols)\protect\cite{Zhu01,Gen}.
The dashed line is the parametrization \protect\cite{Pl90}.}
\label{fig:fig1ia}
\end{figure}

In Fig. \ref{fig:fig1ia} we illustrate the behavior of the different nucleon
electric form factors: \ges,\  \gep\  and \gen. The solid line represents the
Gari-Kr\"umpelmann parametrization \cite{G-K}. 
The new \gep\  data, which decrease faster than the dipole function, are also
well
reproduced by the Gari-Kr\"umpelmann parametrization (solid line). The fit of Eq. \ref{eq:eq5} is shown as a dashed line, and it is almost indistinguishable from  \cite{G-K}.
The  electric neutron form factor is calculated
from the isoscalar nucleon form factor, taking for \gep\  the fit, Eq. (\ref {eq:eq5}), based on the new data (open stars). The results for \gen\  are in
very good agreement with the parametrization
\cite{G-K}. These results show that the neutron form factor becomes more
sizeable
than predicted by other parametrizations, often used in the calculations
\cite{Pl90,Galster} (thin dashed line).
Starting from $Q^2\simeq 2$ (GeV/c)$^2$  the form factor \gen\  becomes even
larger than \gep . Let us mention that the more recent 'direct' measurements
\cite{Zhu01,Gen} are in agreement with
the present values.

\section{Conclusions}

We reviewed the situation with  proton, neutron and deuteron electromagnetic form factors, in the space-like region, at the light of the new data from JLab.
The precise data on proton electric form factors essentially deviate from the dipole approximation. This approximation, previously assumed, was in agreement, on one side with a representation of the nucleon as having an exponetial density distribution, and on the other side, with pQCD prediction. From a theoretical point of view, now the situation is certainly more complex and not yet clarified. 

This new data bring a new view on the commonly assumed description of deuteron form factors.
Let us mention that the $\gamma^*\pi^{\pm}\rho^{\mp}$-contribution, which is a
good approximation for the isoscalar
transition $\gamma^*\rightarrow \pi^+\pi^-\pi^0$
($\gamma^*$ is a virtual photon), is typically considered as the main
correction to IA, necessary, in particular, to improve the description of
the SF $A(Q^2)$ \cite{Ar75}. However the relative role of MEC is strongly model
dependent \cite{Bu92} as the coupling constants for meson-NN-vertexes
are not well known and arbitrary form factors are often added \cite{Ad64,VO95}.
It should be pointed out that the $\gamma^*\pi\rho$ vertex is of
magnetic nature and its contribution to $A(Q^2)$ has to be of the same order of
magnitude as the relativistic corrections.
The electric neutron form factor may be essentially larger than what is commonly assumed, and even larger than the proton form factor, for $Q^2\ge$ 2 GeV$^2$.

The forthcoming data on neutron form factors up to $Q^2$=2 (GeV/c)$^2$, \cite{Madey}
will be crucial in this respect. The large sensitivity of the
deuteron structure to the nucleon form factors
shows the necessity to reconsider the role of meson exchange currents and relativistic corrections in the deuteron physics at large momentum transfer.

A good description of the deuteron will have to take into account in a coherent way, not only wave functions and corrections to IA, but also nucleon from factors.

We thank the Hall A - GEp Collaboration, in particular C. F. Perdrisat, V. Punjabi and M. Jones for many interesting discussions. We thank A. I. Malakhoff and V. Penev, as well as all the organizers of RNP2001 for giving us the possibility to present our results in so loving and historical place. We feel  especially honored to participate in this conference, held in memory of the Academician A. M. Baldin, where a large effort has been undertaken to exploit at best the LHE accelerator, that He conceived.

\end{document}